\def\draftversion{false}

\RequirePackage{ifthen}
\ifthenelse{\equal{\draftversion}{true}}{
  \documentclass[aps,prb,10pt,galley,amsmath,amssymb,showpacs,
       longbibliography,superscriptaddress]{revtex4-1}
}{ 
  \documentclass[aps,prb,10pt,twocolumn,amsmath,amssymb,showpacs,
       longbibliography,superscriptaddress]{revtex4-1}}

\usepackage{lineno} 

\usepackage{graphicx}
\usepackage{physics}
\usepackage[usenames,dvipsnames]{color} 
\usepackage{bm} 
\usepackage{hyperref} 
\hypersetup{		  
    colorlinks=true,
    linkcolor=blue,
    filecolor=blue,      
    urlcolor=blue,
    citecolor=blue
}
\usepackage{array}
\usepackage{dsfont}

\usepackage{soul}

\ifthenelse{\equal{\draftversion}{true}}{
  \marginparwidth 2.7in
  \marginparsep 0.5in
  \newcounter{comm} 
  \def\commnext{\stepcounter{comm}}
  \def\commtext{{\bf\color{blue}[\arabic{comm}]}}
  \def\commmar{{\bf\color{blue}[\arabic{comm}]}}
  \def\dvm#1{\commnext\marginpar{\small DV\commmar: #1}\commtext}
  \def\nvm#1{\commnext\marginpar{\small NV\commmar: #1}\commtext}
  \def\jpm#1{\commnext\marginpar{\small JP\commmar: #1}\commtext}
  \def\jwm#1{\commnext\marginpar{\small JW\commmar: #1}\commtext}
  \def\mlab#1{\marginpar{\small\bf #1}}
  
  \def\parsedate #1:20#2#3#4#5#6#7#8\empty{#4#5/#6#7/20#2#3}
  \def\moddate{\expandafter\parsedate\pdffilemoddate{\jobname.tex}\empty}
  
  \newcommand{\eqlab}[1]{\Red{\hbox{\small\;\;[#1]}}\label{eq:#1}}
  
}{
  \def\dvm#1{}
  \def\nvm#1{}
  \def\jwm#1{}
  \def\jpm#1{}
  \def\mlab#1{}
  
  \newcommand{\eqlab}[1]{\label{eq:#1}}
  
}

\newcommand{\fref}[1]{Fig.~{\ref{fig:#1}}}

\newcommand{\Fref}[1]{Figure~{\ref{fig:#1}}}

\newcommand{\beq}{\begin{equation}}
\newcommand{\eeq}{\end{equation}}
\newcommand{\bea}{\begin{eqnarray}}
\newcommand{\eea}{\end{eqnarray}}
\newcommand{\eq}[1]{Eq.~({\ref{eq:#1}})}

\def\a{{\bf a}}
\def\b{{\bf b}}
\def\c{{\bf c}}
\def\d{{\bf d}}
\def\e{{\bf e}}
\def\f{{\bf f}}

\newcommand{\ignore}[1]{}

\def\zhat{\hat{\bf z}}
\def\nhat{\hat{\bf n}}

\def\k{{\bf k}}

\def\R{{\bf R}}

\def\dag{^\dagger}
\def\pdag{^{\phantom{\dagger}}}

\usepackage[normalem]{ulem}

\begin{document}


\title{Controlling a quantum point junction on the surface of an antiferromagnetic topological insulator}

\author{Nicodemos Varnava}
\email[]{nvarnava@physics.rutgers.edu}
\affiliation{
Department of Physics \& Astronomy, Center for Materials Theory, Rutgers University,
Piscataway, New Jersey 08854, USA}

\author{Justin H. Wilson}
\affiliation{
Department of Physics \& Astronomy, Center for Materials Theory, Rutgers University,
Piscataway, New Jersey 08854, USA}

\author{J. H. Pixley}
\affiliation{
Department of Physics \& Astronomy, Center for Materials Theory, Rutgers University,
Piscataway, New Jersey 08854, USA}

\author{David Vanderbilt}
\affiliation{
Department of Physics \& Astronomy, Center for Materials Theory, Rutgers University,
Piscataway, New Jersey 08854, USA}

\begin{abstract}
The abstract notion of topology has led to profound insights into real materials. 
Notably, the surface and edges of topological materials can host physics, such as unidirectional charge or spin transport, that is unavailable in isolated one- and two-dimensional systems.
However, to fully control the mixing and interference of edge-state wave functions, one needs robust and tunable junctions.
We propose to achieve this control using an antiferromagnetic topological insulator that supports two distinct types of gapless unidirectional channels on its surface, one from antiferromagnetic domain walls and the other from single-height steps.
The distinct geometric nature of these edge modes allows them to intersect robustly to form quantum point junctions, and their presence at the surface makes them subject to control by
magnetic and electrostatic tips like those used in scanning probe microscopes. Prospects for realizing such junctions are encouraged by recent material candidate proposals,
potentially leading to exciting applications in quantum computing and sensing.
\end{abstract}

\maketitle

\section{Introduction}
Surface and edge-state engineering of topological materials
offers great promise for future electronic devices. 
Owing to the topological properties of the bulk of the material, surface states emerge that are protected from elastic and inelastic scattering.
In particular, the community realized early on that topologically-protected chiral (one-way) or helical (two-way) edge states provide dissipationless ``quantum wires'' \cite{hasan-rpm10,qi-prm11} with potential applications in sensor, low-power computing, and quantum information technologies. A crucial part of engineering such wires requires robust and tunable junctions between edge states.

Strikingly, chiral edge states provide directional control of carrier propagation and 
(topological) protection against impurity backscattering.  
This was first demonstrated for quantum Hall edge states in 2D electron gases, but these systems require very low temperature and external magnetic fields. 
A potentially more practical approach to engineering chiral edge states is at the boundary of a 2D quantum anomalous Hall or ``Chern'' insulator.\cite{liu-arcmp16} Experimentally this was first realized in thin films of magnetically doped
topological insulators (TI).\cite{chang-sc13} Unfortunately the inhomogeneity of the magnetic dopants leads
to inevitable disorder\cite{lee-pnas15} and as a result the quantized response
is observed at much lower temperatures than the magnetic gap and Curie temperature allow; to date the state of the art is around $\sim1K$.\cite{checkelsky-nat14,bestwick-prl15,chang-nat15,tokura-nat19} More recently, the discovery of the quantum anomalous Hall effect in twisted bilayer graphene\cite{sharpe-2019,serlin-sc2020} and 5-layer MnBi$_2$Te$_4$\cite{deng-sc20} holds promise for the
realization of topologically protected chiral channels at higher temperatures, due in part to the absence of magnetic-impurity disorder.\cite{wakefield-sc20} 

In its bulk version, MnBi$_2$Te$_4$  belongs to a class of 3D materials that have been variously
described as intrinsic magnetic topological insulators,\cite{gong_cpl19,li-sc19,chen-nat2019,deng-sc20}
axion insulators\cite{wan-prb11,varnava-prb18,varnava-prb20} and second-order topological insulators.\cite{khalaf-prb18,ezawa-prb18,schindler-sa18}
The essential idea is to identify a material whose magnetic symmetry group
enforces\cite{varnava-prb20} a quantized bulk axion coupling\cite{qi-prb08,essin-prl09+e} of $\theta=\pi$,
as in an ordinary 3D TI, but does not enforce the presence of
gapless surface states. Instead, gapped surfaces can appear
naturally on such materials. When they do, they exhibit a
half-quantized surface anomalous Hall conductivity, i.e., an odd integer times $e^2/2h$, 
whose sign is determined by details of
the magnetic order at the terminating surface.  Thus, manipulation
of the surface termination and/or magnetic order in one region
of the surface relative to a neighboring patch, or on one facet
relative to another that meets it at a ``hinge,'' can give rise
to a chiral edge channel at the boundary between these patches
or facets.\cite{varnava-prb18}

In this work we develop a theoretical description for the
creation and 
manipulation of chiral edge
channels on the surface of an antiferromagnetic (AFM) TI.
This class of materials was introduced theoretically by 
Mong and Moore\citep{mong-prb10} and has recently become the focus of intense research with various candidates such as MnBi$_2$Te$_4$\cite{otrokov-nat19}, MnBi$_4$Te$_7$\cite{hu-natcom2020}, $\mathrm{EuIn}_2\mathrm{As}_2$\cite{xu-prl19}
and $\mathrm{NpBi}$\cite{xu-arxiv20} appearing in the  literature. Motivated by these recent developments and the fact
that there is in principle no reason why both the bulk and surface gaps 
could not be on the order of hundreds of meV, allowing for potential
room-temperature device operation, we propose and explore the properties
of a novel \textit{quantum point junction} (QPJ) at the surface of an AFM TI. 

\Fref{intro}$\a,\b$ shows a prototypical spin
arrangement in an AFM TI.  The magnetic
ordering is A-type AFM, i.e., with magnetization uniform in-plane
but alternating from plane to plane along the stacking direction,
which we take to be along $\zhat$. As described in Ref.~\onlinecite{mong-prb10}, each individual layer can be thought of as
adiabatically connected to a 2D Chern insulator, with the sign of
the Chern number alternating from layer to layer.  The sign of the
surface anomalous Hall conductivity of $\pm e^2/2h$ is thus determined by the magnetic
orientation of the last layer at the surface.  As a result, two
kinds of 1D chiral channels can occur at the surface.  As shown
in \fref{intro}$\a$, the emergence of a bulk AFM domain
wall at the surface reverses the sign of the anomalous Hall conductivity on either side of
the resulting line defect, which therefore carries a topologically
protected chiral channel we refer to as a \textit{domain-wall channel}. Alternatively, even if no
bulk AFM domain walls are present, a single-height step can occur
on the surface, as shown in \fref{intro}$\b$.  If it does,
it also marks a sign reversal of the anomalous Hall conductivity when crossing the step,
and thus carries a chiral edge channel as well. We will refer to this as a \textit{step channel}.

\Fref{intro}$\c,\d$, shows the manifestation of the domain-wall and step channel in the surface band structure as described in the context of a tight-binding model used throughout this work (see Methods). The presence of either of these defects results in 1D linear dispersions in the otherwise gapped bulk and surface spectrum of the AFM TI. The states that comprise the chiral bands are exponentially localized in the vicinity of the channel, and host 1D massless Dirac fermions.

\begin{figure}
\centering\includegraphics[width=3.4in]{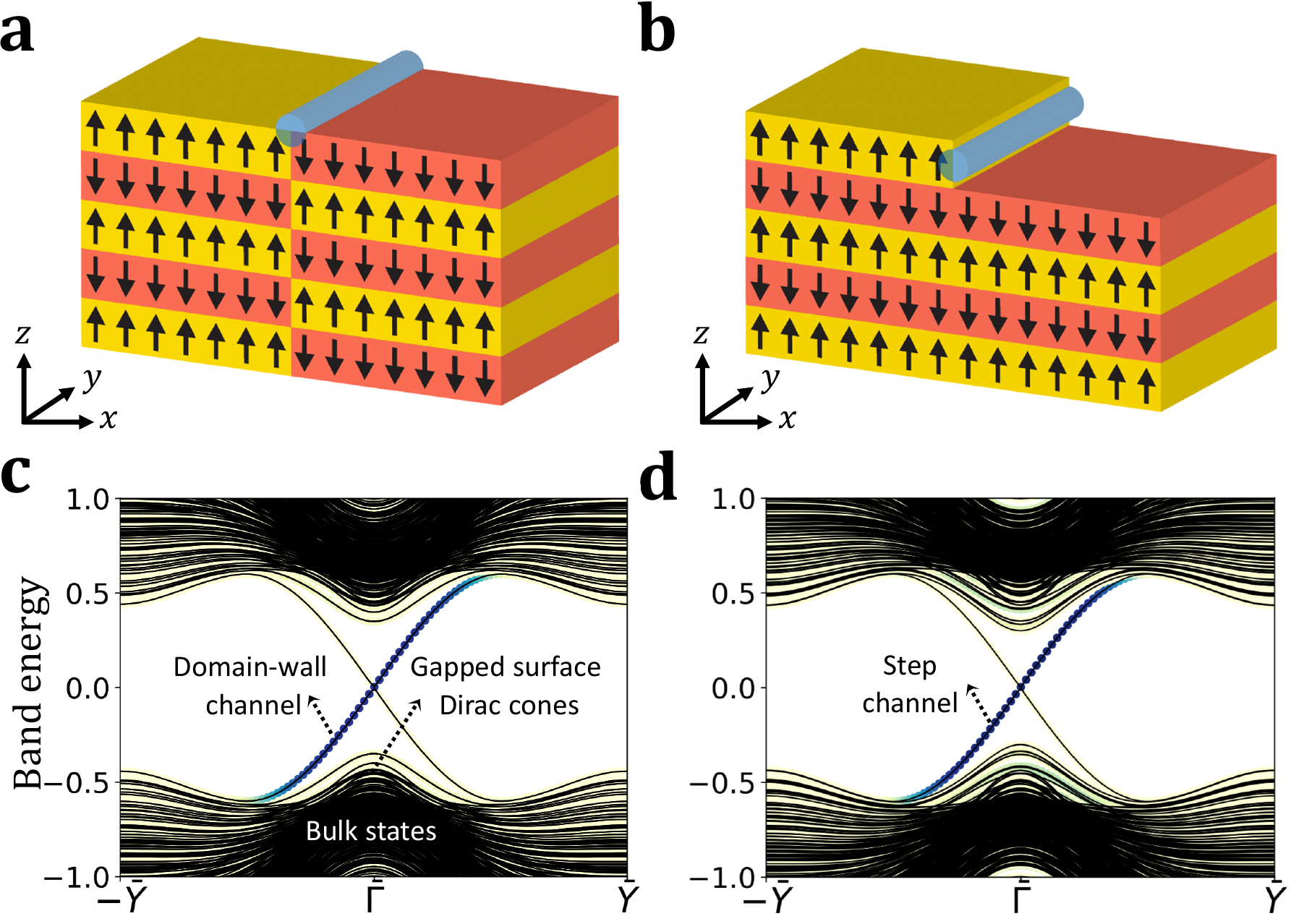}
\caption{$\a,\b$, Depiction of the chiral channel (blue cylinder) at the surface of an A-type AFM TI due to $\a$, a bulk domain wall $\b$, a surface step. $\c,\d$, Surface band structures along (001) in the presence of $\c$, a bulk domain wall $\d$, a surface step. The projection of the states on the chiral channels (blue cylinder) in $\a,\b$, are also shown (blue markers) to illustrate the localization of the massless Dirac fermions that disperse linearly along the channel direction at low energy with velocities $\c$, $v_\text{dw}$  and $\d$, $v_\text{st}$. The description of the model Hamiltonian can be found in Methods. Energies are expressed in terms of the onsite energy $m$ in \eq{afm-ti-model-r}.}
\label{fig:intro}
\end{figure}

The novel opportunity opened by the presence of two different kinds
of 1D chiral channels at the surface, is that these can be
made to intersect, as shown in \fref{junction}$\a$, and such
intersections are expected to remain thermodynamically stable. In contrast, as illustrated in \fref{junction}$\b$,
an intersection between two surface steps can easily evolve via
a pinch-off event into a configuration in which an isthmus of
constant surface height separates the steps; indeed, the width of
such an isthmus will tend to grow due to the line tensions of the
steps, and the quantum junction will have been removed. A similar mechanism affects the intersection of two domain walls.  Instead,
an intersection of a domain wall and step cannot easily be removed,
providing an opportunity for engineering a robust QPJ. In fact, these junctions can appear naturally and were recently seen at the surface of the putative AFM TI MnBi$_2$Te$_4$.\cite{sass-prl20}  Moreover, the fact that this junction occurs at an
exposed surface, not at a buried interface, opens opportunities for
its manipulation by scanning tips of the kind used in scanning
tunneling microscopy (STM) and related methods. Here we are interested in local probes that affect magnetic moments and the electric potential and will refer to them as magnetic and electrostatic STM tips respectively.

In what follows we explore the properties of the QPJ by constructing a tight-binding Hamiltonian associated with the system depicted in \fref{junction}$\a$ and performing dynamic wave-packet (WP)  simulations. Importantly, the creation and detection of single-electron WPs on quantum Hall edge channels has been demonstrated experimentally.\cite{feve-sci07} In fact it has been shown\cite{dubois-nat13,ferraro-prl14,glattli-pssb17} that using modulated bias voltage one can create a minimal excitation known as a Leviton,\cite{levitov-jmp96,ivanov-prb97,keeling-prl06} where the Fermi sea re-organizes itself and a single electron (with no accompanying hole) is emitted in the conduction band. While these techniques directly apply to our proposal, a major benefit of our approach is the formation of thermodynamically stable point junctions which would be difficult to accomplish using quantum Hall edges. In fact, setups like those depicted in the inset of \fref{junction}$\b$, where two chiral channels come in close proximity enabling tunnelling between them, have long been studied in 2D electron gases\cite{wharam-iop88,wees-prl88} and are known as {\it quantum point contacts} (QPC),
a terminology that we have adopted for the junctions proposed here.

As we show below the scattering of WPs at the proposed QPJ can be described by a two level quantum system. In this description the state of the WP is represented by a qubit whose entries correspond to the amplitude (magnitude and phase) of the WP on the two incoming or outgoing channels. Furthermore, the QPJ acts as a unitary quantum gate, or an S-matrix that characterizes the scattering. Remarkably, we find that magnetic and electrostatic STM tips in proximity with the junction can be used to realize any single-qubit gate. In addition, we show that the effect of symmetry breaking terms and weak disorder can be ``gauged away’’ using the two tips. Finally we take a look at potential applications of the QPJ in quantum computing and sensing.

\begin{figure}
\centering\includegraphics[width=3.4in]{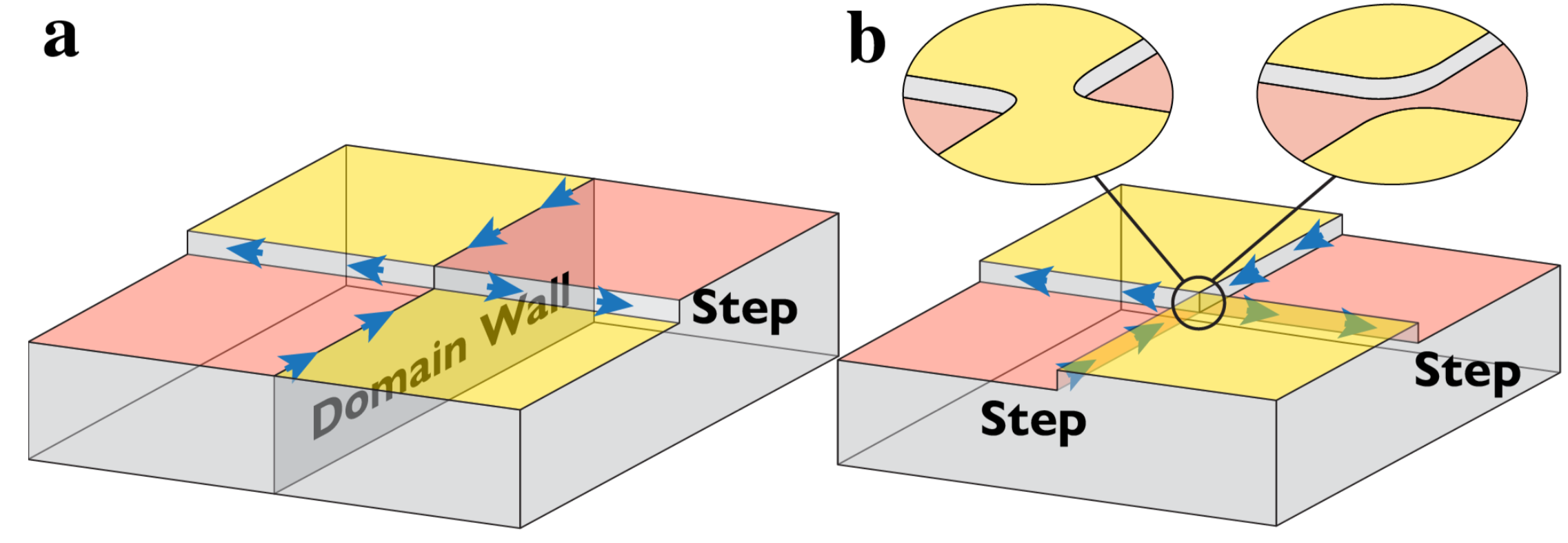}
\caption{$\a$, The intersection of a domain-wall channel with a step channel results in a thermodynamically stable junction, i.e., small surface deformations can only move the junction but not remove it. $\b$, The intersection of two step channels (or two domain-wall channels) is unstable. The inset shows how small deformations remove the junction. Blue arrows indicate the direction of propagation on the chiral channels, while orange and yellow surfaces indicate whether the anomalous Hall conductivity is $\pm e^2/2h$ respectively.}
\label{fig:junction}
\end{figure}

\section{Extracting the $S$ matrix}

\begin{figure*}
\centering\includegraphics[width=7.0in]{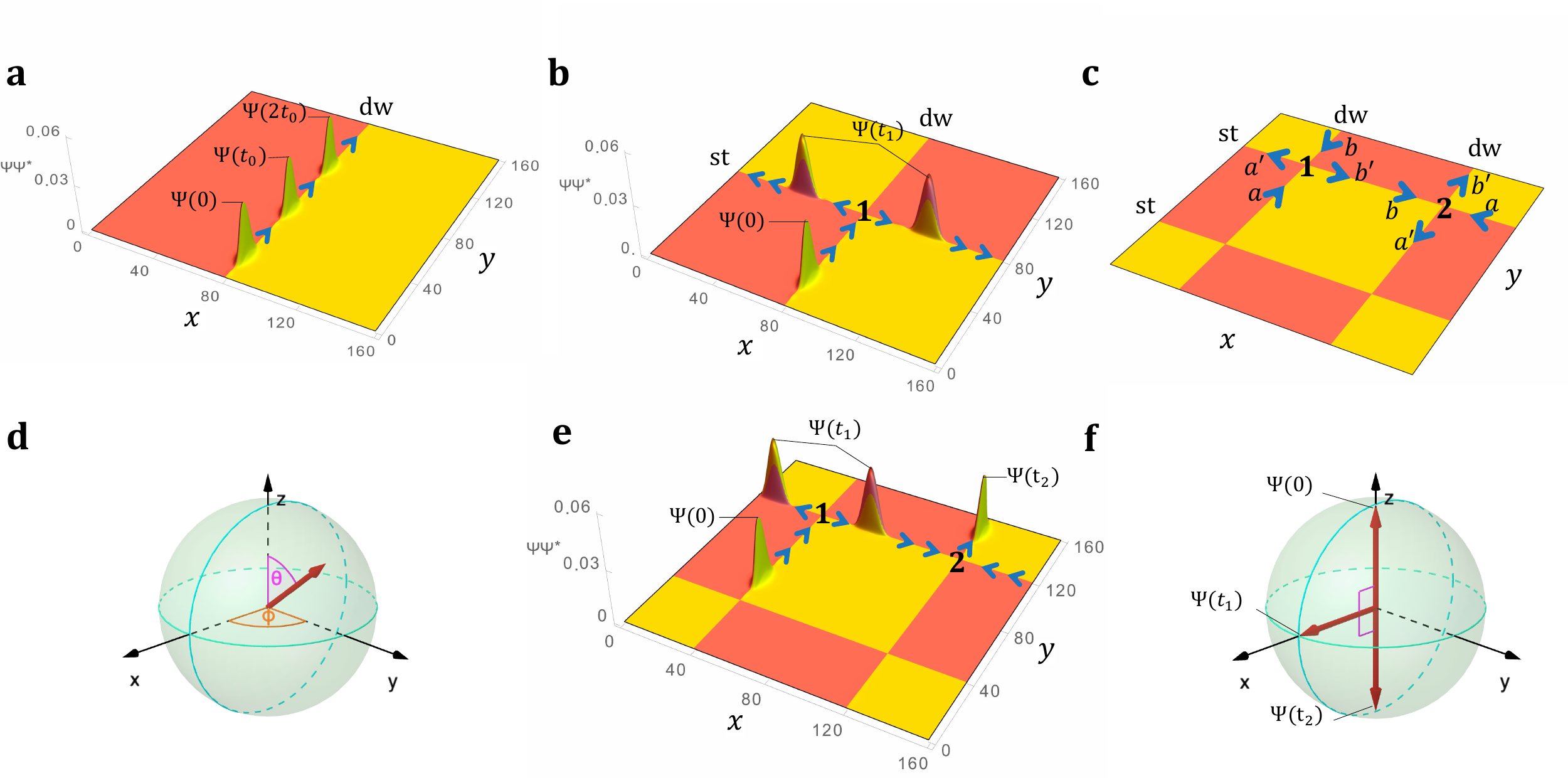}
\caption{
$\a$, Snapshots of the WP showing the propagation on the domain-wall channel.
$\b$, A WP scatters at Junction 1 and splits into two spatially separated outgoing components of the wavefunction. $\c$, Channel labeling convention for Junction 1 and 2. $\d$, Qubit representation of a WP state on the Bloch sphere. $\e$, The initial WP splits after encountering Junction 1, the two components then meet at Junction 2, interfering destructively on channel $a'$ and constructively on channel $b'$. $\f$, Qubit representation of the time evolution in $\e$. WP plots in $\a,\b,\e,$ are calculated from the (001)-projected probability densities at different times. Time-evolution animations showing the WP dynamics in $\a,\e$ are provided in the Supplemental Information.}
\label{fig:setup}
\end{figure*}

We begin by considering the WP dynamics at the surface of an AFM TI. \Fref{setup}$\a$ shows the calculated time evolution of a WP on a single domain-wall channel, while \fref{setup}$\b$, that of a WP in the presence of the QPJ in \fref{junction}$\a$ (see also Supplementary Information for time-evolution animations). In both cases the dissipationless channels are protected from back-scattering by the insulating bulk and surface gaps. The wave function of the WP is thus exponentially confined to the vicinity of the one-dimensional channel, and it travels with a constant group velocity along the channel. In \fref{setup}$\b$, a WP enters along the domain-wall channel, gets split by the QPJ, and then the two components travel away from the QPJ along the step channels. Later we shall consider configurations in which multiple consecutive scattering events occur.

To understand the behaviors observed above, we note that the
wave function of a WP propagating along a single domain-wall channel
in direction $y$, as in \fref{setup}$\a$, can be well approximated\cite{sup-inf} as
\beq
\Psi^{\rm dw}_{\sigma\tau}(x,y,z,t) = \chi^{\rm dw}_{\sigma\tau}(x,z)
   f(y-y_0-v_{\rm dw}t) \,.
\eqlab{wpdw}
\eeq
Here $\chi^{\rm dw}_{\sigma\tau}(x,z)$
captures the transverse shape ($x,z$) and spin-orbital character ($\sigma,\tau$ indices respectively) of the WP\cite{sup-inf},
while $f(y)$ is the envelope function of the WP, which we take to be a Gaussian.
The WP is launched from position $y_0$ at time $t\!=\!0$ and travels with
group velocity $v_{\rm dw}$ (which is set by the surface state dispersion in \fref{intro}$\c$).  In modeling at this level we neglect
spreading of the WP, which we find to be negligible in our simulations.
Similar considerations apply to the propagation of a WP on a step
along $x$ traveling with group velocity $v_\text{st}$ (that is set by the surface state dispersion in \fref{intro}$\d$).

We now consider the scattering event depicted in \fref{setup}$\b$, where an incoming WP splits after encountering the QPJ. We will use unprimed
labels $a$ and $b$ to refer to the two incoming domain-wall channels of Junction 1, as in \fref{setup}$\c$. Note that the extra junctions are the result of in-plane periodic boundary conditions. The incoming initial conditions are
specified by amplitudes
$\phi_a=1$ and $\phi_b=0$.  Now let $t_1$ indicate a time after the
scattering through Junction 1 is complete, but before Junction 2 is encountered.  We label the two outgoing step channels
as $a'$ and $b'$, adopting once and for all the arbitrary convention
that $a\rightarrow a'$ and $b\rightarrow b'$ result from taking
left turns, as shown in \fref{setup}$\c$.
As illustrated in \fref{setup}$\b$, one component of the WP moves to the right and the other to the left, with velocities $v_{\rm st}$ and $-v_{\rm st}$ respectively. At time $t_1$ both will be centered at a distance $x_1$ relative to the junction, so in general we expect to find
\begin{align}
\Psi^{{\rm st}}_{\sigma\tau}(x,y,z,t_1) &=
  \phi_{a'} \tilde{\chi}^{\rm st}_{\sigma\tau}(y,z) f(x+x_1)
\nonumber\\ & \;
 +\phi_{b'} \chi^{\rm st}_{\sigma\tau}(y,z) f(x-x_1) \,.
\end{align}
Here $\phi_{a'}$ and $\phi_{b'}$ are the amplitudes (magnitude and phase) describing scattering from incoming channel $a$ into channels
$a'$ and $b'$ respectively, and
$\tilde{\chi}^{\rm st}$ is the time-reversed partner of $\chi^{\rm st}$. These expectations are well reproduced in our full numerical calculations which therefore allow us to extract the amplitudes 
$\phi_{a'}$ and $\phi_{b'}$.

Similar calculations, where the incident WP approaches
Junction 1 along the $-\hat{y}$ direction on channel $b$, allow us
to extract the corresponding amplitudes that result for initial
conditions of $\phi_a=0$ and $\phi_b$=1.  Thus, we can model
a combined scattering event via
\beq
\begin{pmatrix}
    \phi_{a'} \\
    \phi_{b'} \\
\end{pmatrix} = S
\begin{pmatrix}
    \phi_a \\
    \phi_b \\
\end{pmatrix}
\eqlab{smat}
\eeq
where the elements of the S-matrix are determined by the four complex amplitudes discussed above.

In this way, the evolution of the system of propagating WPs is
mapped onto that of a two-level quantum system, so that it is enough to restrict $S$ to be an SU(2) matrix.\footnote{More generally, systems that involve many junctions and result in $N>2$ output channels will transform under $S \in \textrm{SU}(N)$. In such cases each junction acts as $S \in U(2) \subset \textrm{SU}(N)$. In this work we are only concerned with systems with $N=2$, so that each junction acts as an SU(2) matrix.}
The characterization of a junction by such an S-matrix is a central element of our theory. It is illustrative to represent the initial or final state as a point on the Bloch sphere,
\beq
\begin{pmatrix}
    \phi_a \\
    \phi_b \\
\end{pmatrix} = 
\begin{pmatrix}
    \cos(\theta/2) \\
    e^{i\phi} \sin(\theta/2) \\
\end{pmatrix},
\eqlab{bloch-sph}
\eeq
where $\theta$ determines the relative WP magnitude on channels
$a$ and $b$ and $\phi$ their phase difference, as illustrated
in \fref{setup}$\d$. 

Each junction scattering event can then be described by the
action of the corresponding junction S-matrix on the spinor
representation of the channel states, regarded as a
\textit{qubit state}, and the result of consecutive QPJ scattering
events, as in \fref{setup}$\e$, corresponds to the action
of consecutive 
\textit{gates} acting on these qubits as illustrated in \fref{setup}$\f$.

\begin{figure*}
\centering\includegraphics[width=7.0in]{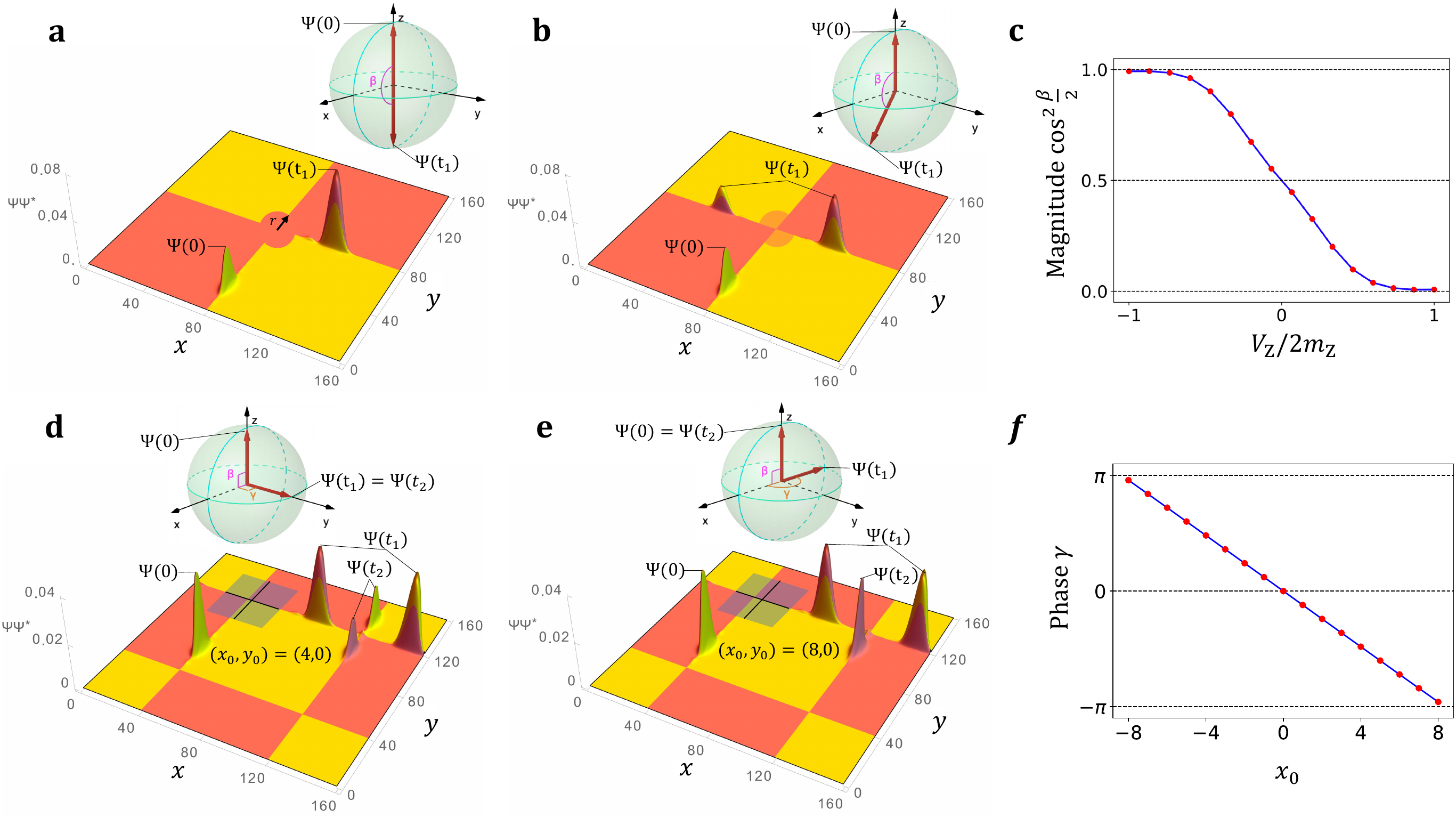}
\caption{Magnitude $\a$-$\c$, and phase $\d$-$\f$, manipulation of $S_1$. $\a$, The magnetic STM tip with $V_\text{Z}=2m_\text{Z}$ in \eq{htip} (see Methods), has polarized the surface spins in a circular region centered at the junction resulting in two uncoupled channels. $\b$, Partially polarized region with $V_\text{Z}=0.4m_\text{Z}$ causes unequal splitting of the WP. $\c$, Numerical calculation of the magnitude splitting $\cos^2{\beta/2}$ as a function of $V_\text{Z}$. Each value corresponds to the integral of $|\Psi(t_1)|^2$ on channel $a'$ of Junction 1. $\d$,$\e$, Applying the electrostatic STM tip with $V_\text{G}=0.6$ in \eq{htip}, at Junction 1 induces a phase difference between the outgoing WPs which then affects how they interfere at Junction 2. $\d$, The center of the rectangular region $\Omega_\text{G}$ is chosen so that $\gamma= \pi/2$. $\e$, Same but we set $ \gamma = \pi$ making the WPs constructively (destructively) interfere on channel $a(b)$ of Junction 2. $\f$, Numerical calculation of the angle $\gamma$ from the relative phase of the outgoing WPs at $t_1$ as a function of $x_0$. The phases of the outgoing WPs are determined from the inner product between $\Psi(t_1)$ in the presence and absence of the phase gate. Time-evolution animations showing the scattering events in $\a,\b,\d,\e$ are provided in the Supplemental Information.}
\label{fig:smat-control}
\end{figure*}

Let us now return to a more specific discussion of our full time-evolution calculations, and our analysis of them in terms of the framework sketched above. \Fref{setup}$\e$, shows the time evolution of a WP initiated on channel $a$. The WP propagates towards and then scatters at Junction 1, splitting into two equal parts. Later the two WPs pass through Junction 2, interfering destructively and constructively on outgoing channels $a'$ and $b'$ respectively. As promised, we can describe the time evolution of the WP configuration as a qubit passing through two gates. Indeed, using the convention of \fref{setup}$\c$, the calculated $S$ matrix of Junction 1 and 2 corresponds to the Hadamard gate 
\beq
S_1=S_2 = \frac{1}{\sqrt{2}}
\begin{pmatrix}
  1 & -1 \\
  1 & \ \  1 \\
\end{pmatrix},
\eqlab{sj-had}
\eeq
so that the final state is related to the initial one by applying the Hadamard gate twice. Geometrically the $S$ matrix expressed as 
\beq
S=R_\mathbf{\hat{n}}(\varphi)=e^{-i\frac{\varphi}{2} \mathbf{\hat{n} \cdot \boldsymbol{\sigma}}} 
\eqlab{rot}
\eeq
describes a qubit rotation by an angle $\varphi$ through an axis $\mathbf{\hat{n}}$ 
and $\boldsymbol{\sigma}=(\sigma_x,\sigma_y,\sigma_z$) is a vector of Pauli matrices.
Since $S_1=S_2=R_\mathbf{\hat{y}}(\pi/2)$, each application rotates the qubit by $90^{\circ}$ around the $\mathbf{\hat{y}}$ axis of the Bloch sphere, resulting in an overall reversal of the pseudospin as shown in \fref{setup}$\f$. 

\section{Controlling the $S$ matrix }

Before explaining how the control of the $S$ matrix is achieved, it is illustrative to break down the action of $S$ into three stages. First we have the propagation along the
incoming channels; since these cannot scatter into one another,
this is represented by a diagonal matrix $S_{\text{dw}}$. Then there
is the scattering $S_{\text{pj}}$ at the QPJ itself, followed
by another channel-diagonal propagation $S_{\text{st}}$ on the
outgoing step channels.  The overall $S$ matrix can then be written in terms
of the Pauli matrices  as
\beq
S= S_{\text{st}} S_{\text{pj}} S_{\text{dw}} = e^{-i\gamma \sigma_z/2} e^{-i\beta \sigma_y/2} e^{-i\alpha \sigma_z/2},
\eqlab{euler}
\eeq
where $S_{\text{pj}}$ is expressed as a real orthogonal matrix because the
phases can be absorbed into $S_{\text{dw}}$ and $S_{\text{st}}$.
Remarkably, $(\alpha,\beta,\gamma)$ are
exactly the three Euler angles that can be used to express
any SU(2) matrix. Thus control over the three Euler angles results in a \textit{universally programmable gate}, which we now demonstrate.

To control the $S$ matrix, we will use two
local probes in the vicinity of the junction. The first one, which we refer to as a magnetic STM tip, affects the local magnetic moments, and as we shall see, controls the magnitudes of the $S$ matrix. The second probe is an electrostatic STM tip modifying the
site energies under the tip, thus controlling the phases of the $S$ matrix. The effect of the magnetic tip is controlled through the coefficient $V_\text{Z}$, and that of the electrostatic tip through $V_\text{G}$, both acting in the local vicinity of the junction. For more details on how this is modeled, see \eq{htip} in Methods.

\paragraph{Magnitude control.}
We set $V_\text{G}=0$, leaving the electric potential constant throughout the crystal so that no extra phase evolution occurs during the propagation ($\alpha=\gamma=0$), and we vary the strength of the magnetic tip $V_\text{Z}$. This  affects the left-right magnitude splitting, i.e., $S = R_\mathbf{\hat{y}}(\beta) = e^{-i\beta\sigma_y/2}$ in \eq{euler}, with $\beta=\beta(V_\text{Z})$. To understand the mechanism behind the magnitude control, first consider the extreme scenario depicted in \fref{smat-control}$\a$. Here a strong magnetic STM tip has polarized the surface magnetization in the vicinity of the junction (orange circular region), forcing the anomalous Hall conductivity to be uniformly $+e^2/2h$ in that area (see Methods). This essentially ``removes'' the junction, and the WP is completely transferred from the domain-wall (channel $a$) to the edge of the step (channel $b'$), so that $S_{1}=R_\mathbf{\hat{y}}(\pi)$. An example of partial polarization, is shown in \fref{smat-control}$\b$, while the results of tuning $V_Z$ over the entire range of tip strength is shown in \fref{smat-control}$\c$, where we plot the numerically calculated value of $\cos^2(\beta/2)$, which represents the asymmetry between left- and right-scattered WPs, as a function of $V_\text{Z}$.  This demonstrates the universal control of the Euler angle $\beta$ using a magnetic STM tip.

\paragraph{Phase control.}
To illustrate the phase control, we set $V_\text{Z}=0$, fix $V_\text{G}$ to a non-zero value (see \eq{htip} in Methods), and control the position of the electrostatic tip. Then $S_j=R_\mathbf{\hat{z}}(\gamma)R_\mathbf{\hat{y}}(\pi/2)R_\mathbf{\hat{z}}(\alpha)$, where $\alpha$ and $\gamma$ are determined by the position $(x_0,y_0)$ of the tip relative to the junction, as described by \eq{relation} (see Methods). The electrostatic tip is depicted as a shaded square with origin $(x_0,y_0)$ in \fref{smat-control}$\d$,$\e$.  In fact, our choice of $\phi_a=1$ and $\phi_b=0$ simplifies the situation, since $R_\mathbf{\hat{z}}(\alpha)$ just corresponds to an overall phase, which is not of interest.  Physically, the WP splits equally at the first junction ($\beta=\pi/2$), and the electrostatic STM tip, corresponding to the second term in \eq{htip}, is then used to control the relative phases of the outgoing WPs via the $R_\mathbf{\hat{z}}(\gamma)$ term.

In \fref{smat-control}$\d$-$\f$, we illustrate the phase control by applying the electrostatic gate on Junction 1. To see the effects of the phase manipulation, we consider the interference that conveniently occurs when the WPs meet again (due to periodic boundary conditions in $x$ and $y$) at Junction 2. In \fref{smat-control}$\d$, the electrostatic tip is centered four unit cells to the right at $(x_0,y_0)=(4,0)$, which approximately makes $\gamma = \pi/2$ so that $S_1 = R_\mathbf{\hat{z}}(\pi/2)R_\mathbf{\hat{y}}(\pi/2)$, while $S_{2} = R_\mathbf{\hat{y}}(\pi/2)$ as before. After scattering at Junction 1 the outgoing WP, whose state corresponds to a vector pointing along the $+y$ direction of the Bloch sphere, becomes the incoming WP at Junction 2. Since Junction 2 acts as a rotation around the $y$-axis, it does not affect the qubit state of the WP.
Similarly, in \fref{smat-control}$\e$, we set $(x_0,y_0)=(8,0)$, so that after encountering Junction 1 the qubit state points along $-\mathbf{\hat{x}}$, and after Junction 2 it returns to its initial $+\mathbf{\hat{z}}$ state.
In \fref{smat-control}$\f$, we present a numerical calculation of $\gamma$ versus $x_0$. This is done by calculating the phase of the WPs just after it scatters off Junction 1. We find a linear behavior as expected from \eq{relation}.

In summary, using the two STM tips we can control $\alpha$, $\beta$, and $\gamma$ independently in \eq{euler}, so that the junction can be made to implement any SU(2) gate.

\paragraph{Symmetries.}
Our QPJ has many artificial symmetries that in any real application will be absent. For example, we have seen that it naturally implements the Hadamard gate so that the WP splits equally after encountering the QPJ; this behavior is enforced by the mirror symmetries $M_x$ and $M_y$ of the bulk Hamiltonian, and should not be expected in general.
We demonstrate in the Supplemental Information that breaking these symmetries does not affect the ability of our protocol to control the $S$ matrix.

\paragraph{Stability to disorder.}
A significant advantage of the QPJ design presented here is that the chiral channels on the domain walls and steps cannot back-scatter and are therefore expected to be robust against the presence of weak disorder (i.e, such that the average bulk and surface gaps remains open). We demonstrate this topological protection of the QPJ by introducing disorder into our model via a short-ranged random potential that is sampled from a Gaussian distribution. Although the qubit gets dephased in a different way for each realization of disorder, as we demonstrate in the Supplemental Information, the electrostatic tip can be used to recalibrate the QPJ. This allows us to remove the random offsets arising from the specific impurity configuration, thus enabling the control of the junction even in the presence of weak disorder.

\section{Applications}

\begin{figure}
\centering\includegraphics[width=\columnwidth]{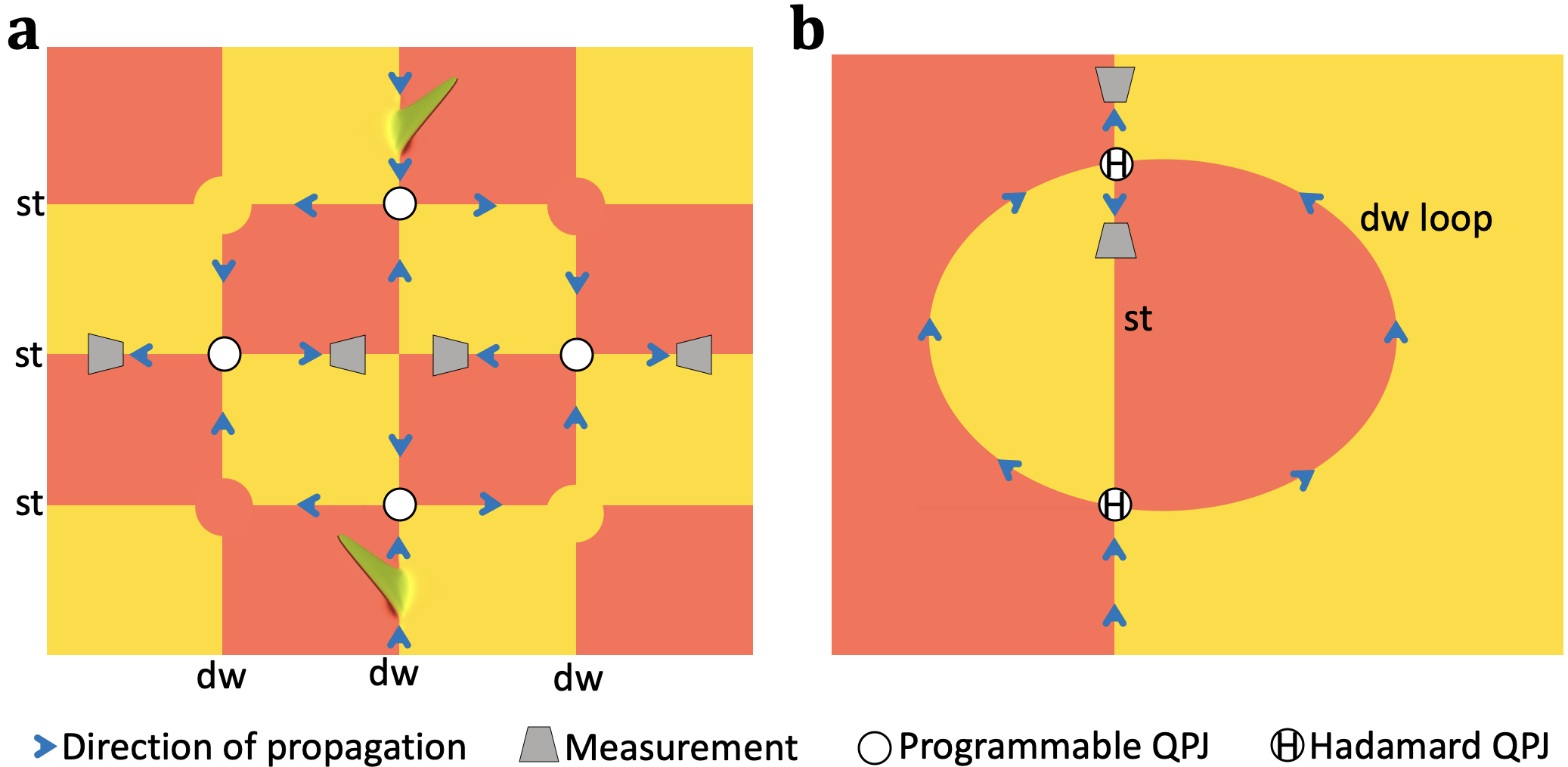}
\caption{Schematic of $\a$, a four-level quantum system with four programmable gates $\b$, a magnetometer consisting of a domain-wall channel loop intersected by a step. }
\label{fig:applications}
\end{figure}

Having theoretically characterized the QPJ, we turn now to a discussion of its potential applications in quantum computing and sensing. 

\paragraph{Quantum computing.} 
Up to this point we have considered systems where an electron WP is injected into the system and after encountering one or more QPJs can be detected at either of two output channels with probabilities given by $|\phi_{a'}|^2,|\phi_{b'}|^2$. More generally one can imagine networks of junctions that implement an $N$-level quantum system, that is $N$ channels where electrons can be detected. An example of a four-level quantum system is shown in \fref{applications}$\a$. Here a network of junctions has been constructed by intersecting parallel domain-wall with parallel step channels. Four of the junctions have been polarised so as to guide the WPs while four others can be programmed. While \fref{applications}{\bf a} uses four junctions to produce a particular four-by-four unitary, universal control over an SU(4) matrix can be achieved with proper placement of six junctions via Givens rotations.\cite{cybenko-cse01} In fact, this can be scaled up to any SU($N$) using at most $N(N-1)/2$ of the fully tunable SU(2) gates we have constructed.

We should note that this approach to quantum-computing belongs to a broader class of approaches associated with \textit{flying-qubit quantum computation}.\cite{ramamoorthy-iop07,yamamoto-nat12}
To date, the discussion has largely focused on the creation and manipulation of WPs on integer quantum-Hall edge states.\cite{bordone-sst19} Importantly, the techniques developed for that approach are transferable to our proposal. A major advantage of our QPJ approach is that it is quite natural for domain walls and steps to intersect, unlike other approaches where there is no obvious mechanism to produce intersections (as in the case of quantum Hall edges).
Furthemore, an AFM TI does not require an external magnetic field to operate, which makes it attractive in microelectronic applications. The time scale of quantum computation is determined by the Fermi velocity of the chiral surface states 
(denoted $\upsilon_{\mathrm{dw}}$ and $\upsilon_{\mathrm{st}}$), 
which typically is of the order  $\hbar\upsilon_\text{F}\sim$ 1eV$\cdot$\AA. Thus, the application of quantum gates is $10^3$ times faster than those of other prominent quantum computation schemes.\cite{haffner-pr08,gambetta-nat17} We also remark that the chiral nature of the surface channels composed of 1D massless Dirac fermions can interface with chiral Majorana fermion qubit architectures, e.g., with the digital qubit gate that has been proposed at quantum anomalous Hall insulator and topological superconductor junctions.\cite{lian-pnas18}

\paragraph{Quantum Sensor:}
Beyond quantum computing, a perhaps simpler application of our QPJ is as a sensitive magnetic sensor. Consider a domain-wall loop channel intersected by a step channel as in \fref{applications}$\b$. In this case we assume a current flowing from the bottom of \fref{applications}$\b$, through the step channel to scatter at the first Hadamard QPJ where it splits equally into two components. The two components then flow on the two domain-wall channels comprising the loop and meet at the second QPJ. If no external magnetic flux threads the loop the two currents will recombine just like in \fref{setup}$\e$, and flow only in the channel that is inside the loop. In contrast, if an external magnetic flux threads the loop the two currents will pick up a relative Aharonov-Bohm phase which affects their interference, resulting in a non-zero current in the channel flowing towards the top of the figure. Thus by measuring the output current one can determine the local magnetic field inside the domain-wall loop region. 

\section{Conclusion}
In this work, we have demonstrated the novel opportunity offered by antiferromagnetic topological insulators to realize robust quantum point junctions at their surfaces, owing to the existence of two types of chiral channels (domain-wall and step) appearing on their surfaces. By identifying the two incoming 
channels  with the two ``code'' states of a qubit, our results show that the $S$ matrix of the quantum point junction acts as a single-qubit gate rotating the state vector of the qubit to produce the two outgoing channels. Furthermore, we have shown that magnetic and electrostatic tips from scanning tunneling microscopy can modify the junction so that its $S$ matrix can perform any rotation on the Bloch sphere, realizing a universal one-qubit gate. In addition, we have considered the effects of symmetries and disorder and have illustrated that these affects can be ``gauged'' away through calibrating the junction. Finally, we have discussed potential applications of the novel quantum point junction as the basic unit to imagine novel devices with applications in quantum computing and sensing. We hope that our approach will inspire new 
paths in exploring complex quantum systems.  

\section{Methods}
\paragraph{Model Hamiltonian.}We consider an adaptation of a simple four-band tight-binding model proposed by Bernevig et al.\cite{bernevig-sci06,hughes-prb11} to describe systems exhibiting a topological phase transition mediated by a single band inversion at $\Gamma$. The simplicity of the model makes detailed calculations practical even for large systems.  The model is written in terms of two spinful orbitals per lattice
site and takes the form
\beq
\begin{split}
H_0 &= m\sum_{\ell} c_\ell\dag \tau^z c\pdag_\ell
+ \frac{t}{2}{\sum_{\ell\ell'}}' c_{\ell}\dag \tau^z c\pdag_{\ell'}  \\
& + \frac{-i\lambda}{2}{\sum_{\ell\ell'}}' c_{\ell}\dag
   \tau^x\nhat\pdag_{\ell\ell'}\cdot \boldsymbol{\sigma}  c\pdag_{\ell'} 
+m_{\text{Z}}\sum_\ell(-)^{\ell_z} c_\ell\dag \sigma^z c\pdag_\ell \;.
\eqlab{afm-ti-model-r}
\end{split}
\eeq
Here $\ell$ labels a lattice site $\R_\ell=(\ell_x,\ell_y,\ell_z)$ on the
 unit cubic lattice, $\sum'_{\ell\ell'}$ indicates a sum over nearest
neighbor sites, and $\nhat\pdag_{\ell\ell'}$ is the nearest neighbor
unit vector.  We have adopted an implied sum notation for the
orbital and spin degrees of freedom, e.g.,
$c_\ell\dag \tau^\mu \sigma^\nu c\pdag_{\ell'} = \sum_{ij,st}
   c^\dagger_{\ell is} \tau^\mu_{ij} \sigma^\nu_{st} c\pdag_{\ell'\!jt}$,
where $\tau$ and $\sigma$ are Pauli matrices for orbital and spin
degrees of freedom respectively, and $c\dag_{\ell is}$ creates an
electron on site $\ell$ in orbital $i$ with spin $s$.

The first three terms in \eq{afm-ti-model-r} correspond to the model of Bernevig et al.\cite{bernevig-sci06,hughes-prb11} for a strong topological
insulator, often written in $k$-space as
$H_{\text{STI}}(\k) = m\tau^z+ \sum_{i=x,y,z} t\cos(k_i)\tau^z
+\lambda\sin(k_i)\tau^x\sigma^i$.  
In the last term in \eq{afm-ti-model-r}, $m_\mathrm{Z}$ is the strength of the staggered
Zeeman field corresponding to A-type (layered) AFM order,
doubling the unit cell and
converting the model to represent an AFM topological insulator.
Time reversal itself is now broken, but time reversal followed by a unit translation along
$\zhat$ is a good symmetry.  For our choice of parameters,\cite{sup-inf} the model
is in the topological phase, with a formal magnetoelectric coupling
of $(\theta/2\pi)(e^2/h)$ with axion coupling $\theta=\pi$.  As
a result, $\zhat$-normal surfaces are naturally gapped and carry
an anomalous Hall conductivity of $\pm e^2/2h$.

\paragraph{Wave-packet construction.} We construct the initial WPs in the space of momentum $k_\parallel$ along the direction of propagation. We calculate the surface band structure for a supercell Hamiltonian $H_{\text{dw}}$ or $H_{\text{st}}$ containing a domain wall or step,
whose presence results in mid-gap bands localized on the conducting channels in the otherwise gapped surface, as shown in \fref{intro}$\c,\d$ respectively. Note that technically each slab contains two domain walls and two steps. In the domain wall case the configuration as a whole is invariant under time reversal times inversion, so the bands shown are Kramers degenerate.

Next we construct the WP by making a quantum superposition of channel-localized solutions according to a $k_\parallel$-space envelope function that we take to be a Gaussian. This results in a WP that is localized in all three real space dimensions.  This is then used as the initial wave function $\Psi(0)$ of the time-evolution problem for the much larger system that includes the QPJ and is described by the Hamiltonian $H_{\text{QPJ}}$. We have defined $H_{\text{QPJ}}$ as the model Hamiltonian $H_0$ in the presence of an antiferromagnetic domain wall and a single-height step that intersect in the center of the surface. A more detailed description of the WP construction can be found in the Supplemental Information. 

\paragraph{Wave-packet dynamics.}
To avoid finite-size effects, we require the system size $L$ to be much larger than the extent of the WPs along the channel. When both a domain wall and step are present, momentum is no longer a good quantum number in any direction, so we compute the time evolution entirely in real space. This is done using Chebyshev series expansion methods~\cite{Weisse-2006} applied to the time-evolution operator $e^{-i Ht}$. We use slabs of size $160\!\times\!160$ in-plane and 16 cells thick, enough to minimize finite-size effects, and adopt a Chebyshev expansion order of $N_C=2^{11}$ so that we can time evolve the state accurately over the needed time intervals.

\paragraph{STM tip modeling.} To model the effects of the magnetic and electrostatic STM tips we extend the QPJ Hamiltonian ($H_{\text{QPJ}}$) with two spatially dependent terms
\beq \tilde{H}_{\text{QPJ}}=H_{\text{QPJ}} + V_\text{Z}\sum_{\ell\in\Omega_\text{Z}} c\dag_\ell \sigma^z c\pdag_\ell + V_\text{G} \sum_{\ell\in\Omega_\text{G}} c\dag_\ell c\pdag_\ell \, ,
\eqlab{htip}
\eeq
where the second term modifies the Zeeman interaction in a region $\Omega_\text{Z}$ and the third term shifts the energy of all orbitals and spins uniformly inside a region $\Omega_\text{G}$.

For a positive $V_\text{Z}$ in \eq{htip}, we choose the region $\Omega_\text{Z}$ such that it restricts the sum to surface orbitals that lie within a radius $r$ of the tip, and that already experience a negative Zeeman field from the bulk Hamiltonian of \eq{afm-ti-model-r}. Thus, $V_\text{Z}=m_\textrm{Z}$ is just enough to remove the Zeeman field from these sites, and $V_\text{Z}=2m_\textrm{Z}$ makes the surface-layer Zeeman field equal on both sides of the domain wall or step, as in \fref{smat-control}$\a$.  We can then tune between these extremes by taking $V_\text{Z} \in [0,2m_\text{Z}]$, thus modeling cases in which the magnetic tip has only partially reversed the surface field. Similarly, for $V_\text{Z}<0$, $\Omega_\text{Z}$ is chosen such that the second term in \eq{htip} is restricted to surface orbitals experiencing a positive Zeeman field in the bulk Hamiltonian.

The region of influence of the electrostatic tip, $\Omega_\text{G}$ in \eq{htip}, is  defined to be a rectangle centered at $(x_0,y_0)$ relative to the QPJ and one unit cell deep, as shown by the grey shading in \fref{smat-control}$\d$.  A WP propagating for a distance $\ell$ along any domain-wall or step channel lying inside the quantum well defined by $\Omega_\text{G}$ acquires an additional phase proportional to $\ell\Delta k$, where $\Delta k$ is the shift of the Fermi wavevector of the channel. In the approximation of linear dispersion, we have $\Delta k=V_\text{G}/\hbar \upsilon_F$,
where $V_\text{G}$ corresponds to a local gate voltage and $\upsilon_F$ is the Fermi velocity 
(equal to $\upsilon_{\text{dw}}$ and $\upsilon_{\text{st}}$ for domain-wall and step channels respectively).
Thus, the off-centering of $\Omega_\text{G}$ defined by $(x_0,y_0)$ allows us to control the travel distances $\ell$ along each of the four ``legs'' near the junction, introducing extra phases that are given by
\beq
\alpha= -\Delta k_{\text{dw}}y_0, \ \gamma=-\Delta k_{\text{st}}x_0
\eqlab{relation}
\eeq
in \eq{euler}.  

\begin{acknowledgments}
This work is supported by NSF CAREER Grant
No. DMR-1941569 (J.H.P.) and NSF Grant DMR-1954856 (N.V. and D.V.).
The authors acknowledge the Beowulf cluster at the Department of Physics and Astronomy of Rutgers University, and the Office of Advanced Research Computing (OARC) at Rutgers, The State University of New Jersey (http://oarc.rutgers.edu), for providing access to the Amarel cluster and associated research computing resources that have contributed to the results reported here.
\end{acknowledgments}

\section*{Author Contributions}
N.V. and D.V. conceived the project. N.V., J.H.W., J.H.P. set up the wave packet simulations that were performed by N.V.; All authors contributed to the theoretical analysis and the writing of the manuscript. \\ 
\section*{Competing interests}
The authors declare no competing interests.

\bibliography{pap}
\end{document}